\documentclass[aps,groupedaddress]{revtex4}
\usepackage{epsfig}
\usepackage{graphicx}
\usepackage[T1]{fontenc}
\usepackage{ae}
\usepackage{ulem}
\usepackage{color}
\usepackage[latin1]{inputenc}
\usepackage{amssymb,amsbsy,amsmath}
\usepackage{bbm}

\newcommand{\op}[1]{\fontdimen12\textfont3=2pt\fontdimen12\scriptfont3=1.4pt\!\null\mathop{\protect\vphantom{#1}\smash{#1}}\limits_{\sim}\null\!}

\newtheorem{prop}{Proposition}
\newtheorem{assu}{Assumption}
\newtheorem{theorem}{Theorem}

\input ulem.sty

\begin{document}


\title[JTE]
      {Stochastic thermodynamics  of a finite quantum system coupled to a heat bath}
\author{Heinz-J\"urgen Schmidt$^1$, J\"urgen Schnack$^2$ amd Jochen Gemmer$^1$ }
\address{$^1$  Universit\"at Osnabr\"uck,
Fachbereich Physik,
 D - 49069 Osnabr\"uck, Germany\\
$^2$  Universit\"at Bielefeld,
   Fakult\"at f\"ur Physik,   D-33501 Bielefeld, Germany}


\begin{abstract}
We consider a situation where an $N$-level system (NLS) is coupled to a heat bath without being necessarily thermalized.
For this situation we derive general Jarzinski-type equations and conclude that heat and entropy is flowing from the
hot bath to the cold NLS and, vice versa, from the hot NLS to the cold bath.
The Clausius relation between increase of entropy and transfer of heat divided by a suitable temperature
assumes the form of two inequalities which have already been considered in the literature. Our approach is illustrated by an analytical example.
\end{abstract}

\maketitle
\section{Introduction}\label{sec:Intro}

The study of non-equilibrium thermodynamics of
systems in contact with a thermal reservoir (``heat bath") of different temperature has a long history.
As an example of the approach via a Master equation and its weak coupling limit we mention the work of Lebowitz and Spohn \cite{SL78},
which actually considers a finite number of reservoirs. During the last decades new methods have been devised,
in particular the approach via fluctuation theorems, see, e.~g., \cite{LS99}.
The famous Jarzynski equation represents one of the rare exact results of nonequilibrium statistical mechanics. It is a statement
about the expectation value of the exponential of the work $\left\langle e^{-\beta \,w}\right\rangle $
performed on a system initially in thermal equilibrium with
inverse temperature $\beta$, but possibly far from equilibrium after the work process. This equation was first formulated for classical
systems \cite{J97} and subsequently proved for quantum systems \cite{K00,T00,M03}. Extensions for systems that are initially in local thermal equilibrium
\cite{T00}, micro-canonical ensembles \cite{TMYH13}, and grand-canonical ensembles
\cite{SS07,SU08,AGMT09,YTC11,CHT11,E12,YKT12} have been published.
The literature on the Jarzynski equation and its applications is rich; a concise review is given in \cite{CHT11},
focusing on the connection with other fluctuation theorems.
The most common approach to the quantum Jarzynski equation is to consider sequential measurements.
This approach is also followed in the present work. The general framework for such an approach was outlined in \cite{SG20a} and \cite{SG20b}.
It is {\it per se} neither quantum mechanical nor classical and will be referred to as ``stochastic thermodynamics'' in the present work.

Interestingly one can derive from the Jarzynski equation certain inequalities that resemble the $2^{nd}$ law, see, e.~g.~, \cite{CHT11}.
However, a closer inspection shows that these inequalities are not exactly statements about the non-decrease of entropy.
But the entropy balance is not the problem:  The total von Neumann entropy is constant during unitary time evolution and  non-decreasing during projective measurements, see \cite{vN32} or \cite[Theorem $11.9$]{NC00}.
The problem is rather that in the quantum case the entropy balance is not sufficient to cover all aspects of the $2^{nd}$ law.

To explain the latter, consider classical thermodynamics where there are several equivalent formulations of the  $2^{nd}$ law.
For example, from the non-decrease in total entropy,
it can be deduced that heat (and entropy) always flows from the hotter to the colder body.
The elementary argument goes as follows: If the hotter body with inverse temperature $\beta$ transfers the infinitesimal heat
$\delta Q$ (of whatever sign) to the colder body with inverse temperature $\beta_0>\beta$ then its entropy decrease will be
$dS_1=\beta \delta Q$, according to the Clausius equality. On the other hand, the colder body receives the heat $\delta Q$ and
its entropy increases by $dS_2=\beta_0 \delta Q$. The total entropy increase will be $dS=dS_2-dS_1= (\beta_0-\beta) \delta Q$.
Since $(\beta_0-\beta)>0$ we obtain $dS\ge 0 \Leftrightarrow \delta Q\ge 0$.

In the quantum case this elementary argument breaks down. Between the two sequential measurements the transferred heat $\Delta Q$
and entropy $\Delta S$ cannot longer be considered as infinitesimal and the Clausius equality has to be replaced by two inequalities.
Moreover, the energy increase of the first system is not exactly equal to the energy decrease of the second system. This holds
only approximately  if the interaction Hamiltonian can be neglected, which will not be the case for small $\Delta Q$.
(Strictly speaking, the latter objection already applies to the classical case.)

We will try here to modify the classical argument for the "correct" heat flow for quantum mechanics.
To this end we will adopt another approach to the problem of the direction of the heat flow that focusses on the $N$-level quantum system
and describes the influence of the heat bath solely in terms of a transition matrix $T$. $T$ is only  a (left) stochastic matrix
and cannot longer assumed to be bi-stochastic (in the strict or modified sense) and hence the usual assumptions leading to
a Jarzynski-type equation, see \cite{SG20a}, are no longer satisfied. But it is possible to derive a more general J-equation that is only based
on (left) stochasticity of $T$. Thus we can find arguments for the ``correct" flow of heat and entropy that only rely on the assumption
that $T$ leaves invariant some Gibbs state with inverse temperature $\beta_0$, see Section \ref{sec:MR}. $\beta_0$
is interpreted as the inverse temperature of the heat bath. In this sense we derive the $2^{nd}$ law of thermodynamics from
a form of the $0^{th}$ law. For this derivation, we combine known results that appear in various places in the literature,
but are sometimes only proved under assumptions that are stronger than those we will assume in the present work.

Jarzynski and W\'{o}jcik \cite{JW04} consider two systems which initially have different temperatures, then interact weakly and finally (in the quantum case) are subjected to a separate energy measurement for both systems. They derive a Crooks-like equality, and from this a Jarzynski-type equation, Eq.~(18) in \cite{JW04},
that corresponds to our Eq.~(\ref{Jar3}). Important assumptions are: Neglect of the interaction between both systems for the heat balance and microreversibility. In contrast, we will focus on  the $1^{st}$ system and describe the $2^{nd}$ one only in a general way by the transition matrix $T$. Our assumptions are weaker: $T$ leaves invariant some Gibbs state without neglecting the interaction; time reversal invariance is not needed (and actually will be violated for the example
presented in Section \ref{sec:AE}).

Jennings and Rudolph \cite{JR10} also consider two systems, which, however, can also be initially entangled. The special case, which is interesting for our purposes, is that both systems are uncorrelated at the beginning and have different temperatures. A first Clausius inequality, corresponding to our (\ref{RCI}), is derived from the property that the Gibbs state minimizes the free energy. From this directly, without using a Jarzynski-type equation, the heat flow inequality, Eq.~(3) in \cite{JR10}, follows, which corresponds to our Eq.~(\ref{JIT2}), but again assuming, as in \cite{JW04}, that the amount of heat emitted by the first system is exactly absorbed by the second one.

The second Clausius inequality that appears in our (\ref{theorem4}), can also be found in \cite[Chapter 4.1]{Sa20},  and is proved there via
the ``monotonicity of the Kullback-Leibler (KL) divergence".
This proof is closely related to ours, since the said monotonicity follows from the general J-equation, see Appendix \ref{sec:GJE}.
The following statements in \cite{Sa20} interpreting the second Clausius inequality as a form of the $2^{nd}$
law of stochastic thermodynamics should be taken with some caution, see the discussion above.

Another related result has been proven already in 1978 by Spohn \cite{S78}: For an open quantum mechanical system described
by a quantum dynamical semigroup that leaves invariant a certain state $\rho^0$ the {\it entropy production} is non-negative.
The latter is defined as the time derivative of the KL divergence between $\rho_t$ and $\rho^0$. This result has been recently
reformulated in \cite{AMH21} in a way compatible with our approach;
Eqs.~(16) and (17) of that reference immediately imply the second Clausius inequality.

Summarizing the state of research, partial formulations of the $2^{nd}$ law for the coupling of an NLS to a heat bath can be found to a sufficient extent in recent years, but they have to be integrated into a unified theory and proved under conditions as weak as possible. This will be attempted in the present work.

The structure of the paper is as follows. The general definitions and main results on the heat flow between the system and the heat bath are given in Section \ref{sec:MR}. These results are based on a Jarzynski-type equation (\ref{Jar1}), which is proved in a more general setting in the appendix \ref{sec:GJE}.
Here we also comment on the possibility to relax the usual assumption of an initial product state.
Close to the equilibrium point the entropy increase $\Delta S$ as well as the absorbed heat over temperature $\beta \Delta Q$ have a common
tangent, see Figure \ref{FIGDES}, with a positive slope, as proved in the Appendix \ref{sec:PR1}. The analogous result on the entropy flow
is formulated in Section \ref{sec:CI}. It depends on two ``Clausius inequalities", see Eq.~(\ref{theorem4}), the second one of which again
follows from the Jarzynski-type equation. The bi-stochastic limit is shortly considered in Section \ref{sec:BL}. The next Section \ref{sec:AE}
contains an analytically solvable example. We close with a summary and outlook in Section \ref{sec:SO}.

\section{Main results on heat flow}\label{sec:MR}

We consider an $N$-level system (NLS) described by a finite index set ${\mathcal N}$,
energies $E_n$ and degeneracies $d_n$ for $n\in{\mathcal N}$. The NLS is assumed to be initially in a Gibbs state
with probabilities
\begin{equation}\label{defpi}
  p_n = \frac{d_n}{Z}\,\exp\left(-\beta\,E_n \right)
  \;,
\end{equation}
where the partition function $Z$ is defined by
\begin{equation}\label{defZ}
  Z =\sum_n d_n\,\exp\left(-\beta\,E_n \right)
  \;,
\end{equation}
and $\beta=\frac{1}{\tau}$ is the inverse temperature of the NLS.
After an interaction with a heat bath a subsequent measurement of energy finds the NLS in the level $m\in{\mathcal N}$  with probability
\begin{equation}\label{defqm}
 q_m= \sum_n P(m\leftarrow n)\,p_n \equiv \sum_nT_{m n}\,p_n
 \;.
\end{equation}
Here the ``transition matrix" $T$ is an $N\times N$ (left) stochastic matrix, i,~e., satisfying
\begin{eqnarray}
\label{Trans1}
 T_{mn }&\ge& 0 \quad \mbox{for all } m,n \in{\mathcal N}\;, \\
 \label{Trans2}
  \sum_m T_{mn} &=&1 \quad \mbox{for all } n \in{\mathcal N}\;.
\end{eqnarray}
The entries of $T$ will be sometimes written as conditional probabilities $T_{mn}=P(m\leftarrow n)$ with self-explaning notation.
We do not make any assumptions concerning thermalization and hence the final probabilities $q_m$ will, in general, not be of Gibbs type.

The usual Jarzynski equation is based on the property of $T$ being a bi-stochastic matrix, i.~e., additionally satisfying,
in the non-degenerate case of $d_n\equiv 1$,
\begin{equation}\label{bistoch}
  \sum_n T_{mn}=1\quad \mbox{for all } m \in{\mathcal N}\;,
\end{equation}
or, in the general case,
\begin{equation}\label{bistochmod}
  \sum_n T_{mn} d_n=d_m\quad \mbox{for all } m \in{\mathcal N}\;,
\end{equation}
see Eq.~(25) in \cite{SG20a}. This property holds if the system is closed and only subject to external forces performing work
upon the system. But bi-stochasticity is no longer guaranteed for systems coupled to other ones (heat baths).
However, if this is the case and if no external forces are applied, we may relax the (modified) bi-stochasticity of $T$
to the following:
\begin{assu}\label{A1}
There exists a Gibbs state with probabilities
\begin{equation}\label{gibbsA1}
 p^{(0)}_n=\frac{d_n}{Z_0}\,\exp\left( -\beta_0 \, E_n\right)
 \;,
 \end{equation}
and
\begin{equation}\label{ZA1}
 Z_0= \sum _n d_n\,\exp\left( -\beta_0 \, E_n\right)
 \;,
 \end{equation}
that is left fixed by $T$, i.~e.,
\begin{equation}\label{TA1}
  \sum_n T_{mn}\,p^{(0)}_n = p^{(0)}_m \quad \mbox{for all } m \in{\mathcal N}\;.
\end{equation}
\end{assu}

In this case the transition matrix $T$, satisfying (\ref{Trans1}), (\ref{Trans2}) and (\ref{TA1}),
will be called a ``Gibbs matrix" with temperature $\tau_0=\frac{1}{\beta_0}$. We will also refer to $\tau_0$
as the ``temperature of the heat bath".

Recall that every (left) stochastic matrix $T$ has an eigenvector $p^{(0)}$ with non-negative entries corresponding to the eigenvalue $1$,
although $p^{(0)}$  is generally not unique. If the entries $p^{(0)}_n$ are different and positive, one can always define
suitable energies $E_n:=-\log p^{(0)}_n$ such that (\ref{gibbsA1}) holds with $\beta_0=1$ and $d_n\equiv 1$.
This has been used to generate numerical examples of Gibbs matrices, see Figures \ref{FIGDES} and \ref{FIGDES1}.
Mathematically, $T$ being bi-stochastic  is a special case of being a Gibbs matrix,
since (\ref{bistoch}) follows from (\ref{TA1}) for $\beta_0=0$ and $d_n=1$ for all $n\in{\mathcal N}$.
However, according to the above remarks, being a Gibbs matrix should rather be considered as a property
of $T$ relative to a given family of energies $E_n$, not as a property of $T$ alone.

Physically, the property (\ref{TA1}) appears plausible if $T$ represents the transition matrix due to the interaction
with a heat bath of temperature $\tau_0$. If the NLS has already the same temperature $\tau_0$ its state should not change.
This is not trivial since, in general, the Gibbs state of the combined system, NLS plus heat bath, with temperature $\tau_0$
does not commute with the interaction Hamiltonian. However, it can be shown that (\ref{TA1})
holds exactly for some analytically solvable examples \cite{SS21},
and in other cases the real situation can be expected to be represented by (\ref{TA1}) to an excellent approximation.

Next, we will recall some probabilistic framework concepts for the Jarzynski-type equation, see \cite{SG20a}.
Let ${\mathcal N}\times {\mathcal N}$ be the set of ``elementary events" such that one event
$(m,n)\in {\mathcal N}\times {\mathcal N}$ represents the outcome of a sequential energy measurement
at the NLS in the sense that the initial measurement yields the result $E_n$, and, after the interaction
with the heat bath, the final measurement yields $E_m$. The probability function
\begin{equation}\label{probfunc}
 P:{\mathcal N}\times {\mathcal N}\longrightarrow [0,1]
\end{equation}
defined for elementary events is given by
\begin{equation}\label{pmn}
 P(m,n)= P(m\leftarrow n)\,p_n = T_{mn}\, \frac{d_n}{Z}\,\exp\left( -\beta E_n\right)
 \;.
\end{equation}
Analogously to the case of the ordinary Jarzynski equation we consider
random variables $Y:{\mathcal N}\times {\mathcal N}\longrightarrow \mathbbm{R}$
and their expectation value denoted by
\begin{equation}\label{defexpY}
\langle Y \rangle := \sum_{mn}P(m,n)\,Y(m,n)= \sum_{mn}T_{mn}\,p_n\,Y(m,n)
\;.
\end{equation}
An example is
\begin{equation}\label{deltaQ}
  \Delta Q: {\mathcal N}\times {\mathcal N}\longrightarrow {\mathbbm R}
\end{equation}
defined by
\begin{equation}\label{defdeltaQ}
 \Delta Q(m,n) := E_m -E_n
 \;.
\end{equation}
This can be interpreted as the ``heat" transferred to the NLS during the interaction with the heat bath
since we have assumed that no external forces are active that could perform work on the NLS.\\

Another example is the random variable ``entropy increase"
\begin{equation}\label{deltaS}
  \Delta S: {\mathcal N}\times {\mathcal N}\longrightarrow {\mathbbm R}
\end{equation}
defined by
\begin{equation}\label{defdeltaS}
 \Delta S(m,n) :=  \log \frac{p_n}{d_n} - \log \frac{q_m}{d_m}
 \;.
\end{equation}
To show the consistency of the definition (\ref{defdeltaS}) we calculate its expectation value
\begin{eqnarray}
\label{expDelta S1}
  \langle \Delta S\rangle  &\stackrel{(\ref{defexpY})}{=}& \sum_{mn}T_{mn}\,p_n\, \left(  \log \frac{p_n}{d_n} - \log \frac{q_m}{d_m}  \right) \\
   \label{expDelta S2}
    &\stackrel{(\ref{defqm},\ref{Trans2})}{=}& \sum_{n}p_n\, \log \frac{p_n}{d_n} -\sum_{m}q_m\,\log \frac{q_m}{d_m}\\
    \label{expDelta S3}
    &=:& S(q)-S(p)
    \;,
\end{eqnarray}
anticipating the definitions (\ref{defSp}), (\ref{defSq}) of the next Section \ref{sec:CI}.

Sometimes, instead of (\ref{defexpY}), we will also use the sloppy notation $\langle Y(m,n)\rangle$ for the expectation value.\\
Then we can state the following Jarzynski-type equation that follows from the general "J-equation" considered in Appendix \ref{sec:GJE}.
\begin{theorem}\label{T1}
If $T$ is a Gibbs matrix with inverse temperature $\beta_0$ and
$\widetilde{p}$ an arbitrary probability distribution, hence satisfying
\begin{equation}\label{sumtildep}
  \sum_{n\in {\mathcal N}} \widetilde{p}_n =1
  \;,
\end{equation}
then, under the preceding conditions,
the following holds:
\begin{equation}\label{Jar1}
 \left\langle \frac{p^{(0)}_n\,\widetilde{p}_m}{p^{(0)}_m\,{p}_n}\right\rangle=1
 \;.
\end{equation}
\end{theorem}
For the proof see Appendix \ref{sec:GJE}, where (\ref{Jar1}) is obtained as a special case.

The Jarzynski-type equation (\ref{Jar1}) is more of a template that can be used to generate
further equations by choosing a special form of the general probability distribution $\widetilde{p}$.
As a particular choice we will consider $\widetilde{p}=p$. This yields
\begin{equation}\label{Jar2}
 \left\langle \frac{p^{(0)}_n\,p_m}{p^{(0)}_m\,{p}_n}\right\rangle=1
 \;,
\end{equation}
and further, using
\begin{eqnarray}
\label{pqu1}
  \frac{p_m}{p_n} &=& \frac{d_m}{d_n}\exp\left(-\beta \left(E_m-E_n\right) \right)\;, \\
  \label{pqu2}
   \frac{p^{(0)}_n}{p^{(0)}_m} &=& \frac{d_n}{d_m}\exp\left(-\beta_0 \left(E_n-E_m \right)\right)
   \;,
\end{eqnarray}
and (\ref{defdeltaQ}), the following equation:
\begin{equation}\label{Jar3}
 \left\langle {\sf e}^{-\left( \beta-\beta_0\right) \Delta Q}\right\rangle=1
 \;.
\end{equation}
This equation was also derived in \cite{JW04} under stronger assumptions. As pointed out in \cite{JW04},
Eq.~(\ref{Jar3}) implies that the probability of events where heat flows in the ``wrong" direction, i.~e., where
$\left( \beta-\beta_0\right) \Delta Q <0$, must be exponentially suppressed.
The reason is that in the case of a Jarzynski-type equation of the form $\left\langle {\sf e}^X\right\rangle=1$
the contributions to the expectation value from large positive values of $X$ must be
counterbalanced by a large number of contributions from negative values of $X$ in order to maintain the expectation value at $1$.

As for the original Jarzynski equation we may derive an inequality by invoking Jensen's inequality (JI).
Note that $x\mapsto - \log x$ is a convex function. Hence
\begin{eqnarray}
\label{JI1}
  0 &=&- \log 1 \stackrel{(\ref{Jar3})}{=} -\log \left\langle \exp\left(-\left( \beta-\beta_0\right) \Delta Q \right)\right\rangle  \\
   &\stackrel{(JI)}{\le}& \left\langle  -\log \exp\left(-\left( \beta-\beta_0\right) \Delta Q \right)\right\rangle
   = \left(\beta-\beta_0\right) \, \left\langle \Delta Q \right\rangle
   \;.
\end{eqnarray}
Thus we have proven:
\begin{theorem}\label{T2}
   \begin{equation}\label{JIT2}
    \left(\beta-\beta_0\right) \, \left\langle \Delta Q \right\rangle \ge 0
    \;.
  \end{equation}
\end{theorem}
If the temperature $\tau$ of the NLS is lower than the temperature $\tau_0$ of the heat bath, then
$\beta-\beta_0\ge 0$ and hence, by means of (\ref{JIT2}), $\left\langle \Delta Q \right\rangle \ge 0$.
That means that in this case the expectation value of the heat flowing into the NLS will be positive, and vice versa.
In other words, heat will flow from the hotter body to the colder one, analogously to the result in classical
thermodynamics, see the Introduction.

\section{Clausius inequalities}\label{sec:CI}

We adopt the notation of the preceding Section \ref{sec:MR} but for the next steps will not need  Assumption \ref{A1}.
Further define (sometimes skipping the expectation brackets $\langle \ldots \rangle$ if no misunderstanding can occur):
\begin{eqnarray}
\label{defEp}
  E(p) &:=& \sum_n p_n\,E_n, \\
  \label{defEp2}
   E_2(p) &:=& \sum_n p_n\,E_n^2, \\
    \label{defEq}
  E(q) &:=& \sum_n q_n\,E_n,\\
  \label{defDE}
  \Delta E&:=& E(q)-E(p),\\
  \label{defSp}
  S(p)&:=&- \sum_n p_n\,\log \frac{p_n}{d_n}, \\
  \label{defSq}
   S(q)&:=&- \sum_n q_n\,\log \frac{q_n}{d_n}, \\
   \label{defDS}
    \Delta S &:=& S(q)-S(p)
    \;.
\end{eqnarray}
Note that (\ref{defpi}) implies the familiar identity
\begin{equation}\label{Sqe}
  S(p)\stackrel{(\ref{defpi})}{=} \sum_n p_n \left( \beta E_n+\log Z\right)\stackrel{(\ref{defEp})}{=}\beta E(p) + \log Z
  \;.
\end{equation}
Then we can show the following:
\begin{theorem}\label{T3}
  Under the preceding conditions the ``first Clausius inequality"
  \begin{equation}\label{RCI}
    \beta \langle\Delta E \rangle \ge  \langle\Delta S \rangle
\end{equation}
  holds.
\end{theorem}
This inequality has also be obtained in \cite{JR10} by considering two systems with weak interaction and using the fact that the Gibbs state
minimizes the free energy. This statement, in turn, can also be proven by the Gibbs inequality used below.

{\bf Proof} of Theorem \ref{T3}:
For any two probability distributions $p,q:{\mathcal N}\rightarrow [0,1]$ there holds the {\it Gibbs inequality} or
{\it non-negativity of the KL-divergence}
\begin{equation}\label{GI}
 S(q||p):= \sum_n q_n \log \frac{q_n}{p_n}\ge 0
  \;,
\end{equation}
see, e.~g., \cite[Theorem $11.1$]{NC00}. From this we obtain
\begin{equation}\label{KIpq}
 -S(q)\stackrel{(\ref{defSq})}{=} \sum_n q_n \,\log \frac{q_n}{d_n} \stackrel{(\ref{GI})}{\ge}  \sum_n q_n \,\log \frac{p_n}{d_n}
 \stackrel{(\ref{defpi},\ref{defEq})}{=}-\beta E(q)-\log Z
 \;,
\end{equation}
and further
\begin{eqnarray}
\label{CI1}
 \beta \Delta E &\stackrel{(\ref{defDE})}{=}&\beta \left( E(q)-E(p)\right)\\
  &\stackrel{(\ref{Sqe})}{=}& \beta E(q)-S(p)+\log Z\\
  &\stackrel{(\ref{KIpq})}{\ge}& S(q)-S(p) \stackrel{(\ref{defDS})}{=} \Delta S
  \;,
\end{eqnarray}
which concludes the proof of Theorem \ref{T3}. \hfill$\Box$\\

\begin{figure}[t]
\begin{center}
  \includegraphics[width=0.7\linewidth]{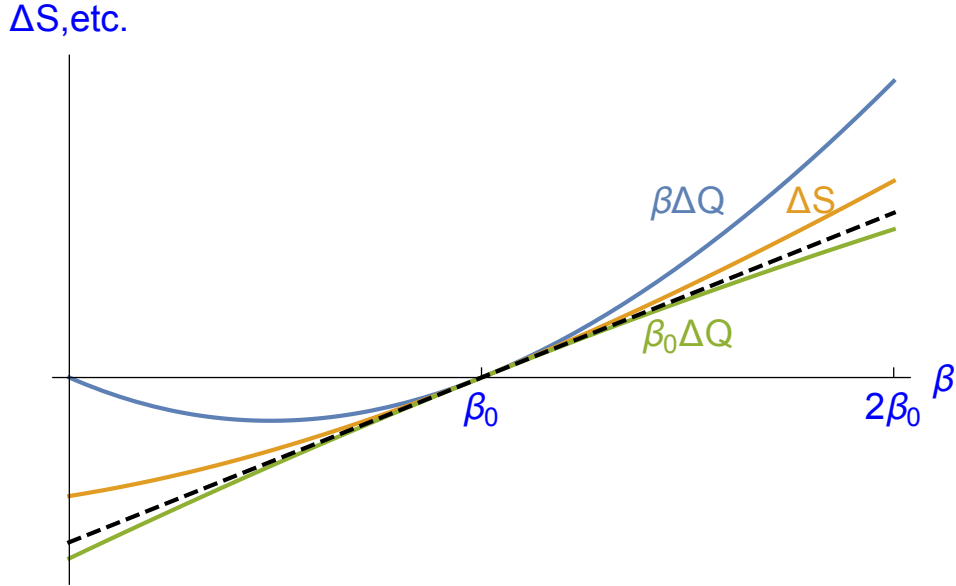}
\end{center}
\caption{Typical plot of the two ``Clausius heat terms" $\beta_0 \Delta Q$, $\beta \Delta Q$ and  the ``entropy increase"
$\Delta S$ as  functions of the inverse temperature $\beta$ of the NLS calculated for a randomly chosen transition matrix $T$ and $N=4$.
Note that $\beta_0 \Delta Q\le \Delta S\le \beta \Delta Q$ holds in accordance with (\ref{theorem4}).
The inverse temperature of the heat bath is $\beta_0$, where all three functions vanish and have a common tangent (dashed black line)
with the slope $a$ according to (\ref{betaDQ0}).
}
\label{FIGDES}
\end{figure}

Interestingly, the first Clausius inequality can be sharpened to a {\it Clausius equality} in the weak coupling limit.
\begin{prop}\label{PCE}
If the transition matrix $T$ is of the form
\begin{equation}\label{Tweak}
 T=\mathbbm{1}+\varepsilon\, t+O\left(\varepsilon^2\right)
 \;,
\end{equation}
where $t$ denotes some $N\times N$-matrix with necessarily vanishing column sums,
then
\begin{equation}\label{DeltaSweak}
  \Delta S =\beta \Delta E +O\left(\varepsilon^2\right)
  \;.
\end{equation}
\end{prop}
The proof can be found in the Appendix \ref{sec:PCE}.\\

Next we assume the situation of a ``heat process" as in Section \ref{sec:MR} together with
Assumption \ref{A1} and hence can interpret the energy difference $\langle\Delta E\rangle$ as the heat $\langle\Delta Q\rangle$
transferred to the NLS.
We consider both sides of (\ref{RCI}), $\beta  \langle\Delta Q  \rangle$ and $ \langle\Delta S  \rangle$,  as functions of the inverse temperature $\beta$.
Both functions vanish at the inverse temperature $\beta_0$ of the heat bath and, due to (\ref{RCI}), must have a common
tangent at $\beta=\beta_0$, see Figure \ref{FIGDES}. We will calculate its slope $a$ using the intermediate results
\begin{eqnarray}
\label{ir1}
  \frac{\partial p_n}{\partial \beta} &\stackrel{(\ref{defpi},\ref{defZ})}{=}& p_n\,\left( E(p)-E_n\right) \\
  \label{ir2}
 \frac{\partial E(p)}{\partial \beta} &\stackrel{(\ref{defEp},\ref{ir1})}{=}&\sum_n p_n\left( E(p)-E_n\right) E_n
 \stackrel{(\ref{defEp2})}{=} E(p)^2-E_2(p)\\
 \label{ir3}
  \frac{\partial q_n}{\partial \beta} & \stackrel{(\ref{defqm})}{=}& \sum_m T_{nm}\frac{\partial p_m}{\partial \beta}
  \stackrel{(\ref{ir1})}{=}\sum_m T_{nm}\,p_m\,\left( E(p)-E_m\right)= q_n\, E(p) -\sum_m T_{nm}\,p_m\,E_m\\
  \label{ir4}
\frac{\partial E(q)}{\partial \beta} &\stackrel{(\ref{defEq})}{=}& \sum_n  \frac{\partial q_n}{\partial \beta}\,E_n
\stackrel{(\ref{ir3})}{=}\sum_n  \left(q_n\, E(p) -\sum_m T_{nm}\,p_m\,E_m\right)E_n
=E(q) E(p)-\sum_{nm}T_{nm}\,p_m\,E_m\,E_n
\;.
\end{eqnarray}
This yields
\begin{eqnarray}\label{betaDQ0}
a&:=&\left. \frac{\partial }{\partial \beta}\beta \langle \Delta Q\rangle\right|_{\beta=\beta_0}\\
\label{betaDQ1}
&=&\underbrace{ \left. \langle\Delta Q\rangle\right|_{\beta=\beta_0}}_{=0}+\beta_0 \,
\left. \frac{\partial }{\partial \beta}\langle\Delta Q\rangle\right|_{\beta=\beta_0}\\
\label{betaDQ2}
&\stackrel{(\ref{defDE},\ref{ir2},\ref{ir4})}{=}&
\beta_0\left(\left.E(q)E(p)-\sum_{nm}T_{nm}\,p_m\,E_m\,E_n-E(p)^2+E_2(p)\right|_{\beta=\beta_0}\right)\\
\label{betaDQ3}
&=&\beta_0\left(E_2(p^{(0)})-\sum_{nm}T_{nm}\,p^{(0)}_m\,E_m\,E_n\right)\\
\label{betaDQ4}
&=& \beta_0\left(\sum_{nm}T_{nm}\,p^{(0)}_m\,E_m\left( E_m-E_n\right)\right)
\;,
\end{eqnarray}
using Eq.~(\ref{Trans2}) in (\ref{betaDQ4}).
According to Theorem \ref{T1} it is clear that the slope of the tangent cannot be negative, $a\ge 0$.
Nevertheless, this will be checked independently, see Appendix \ref{sec:PR1}.

The linear part of the Taylor series of $\beta \langle\Delta Q \rangle$ w.~r.~t.~$(\beta-\beta_0)$
can be used to re-writing $\langle\Delta Q \rangle$ as a function of the dimensionless temperature
$\tau=\frac{1}{\beta}$ such that the zero of $\langle\Delta Q\rangle$ at $\beta=\beta_0$ corresponds to
the temperature $\tau_0=\frac{1}{\beta_0}$. The result
\begin{equation}\label{linQ}
 \langle\Delta Q \rangle = -a\,\beta_0\,\left(\tau -\tau _0\right)+ O\left(\tau -\tau _0\right)^2
\end{equation}
resembles the Fourier law or its precursor, Newton's law of cooling \cite{N01},
stating that the rate of heat loss of a body is directly proportional to the difference in the temperatures between the body and its surroundings.
We further remark that the explicit form (\ref{betaDQ4}) of the ``heat conduction coefficient" $a\,\beta_0$ in (\ref{linQ}) is reminiscent of the fluctuation-dissipation theorems mentioned  in \cite{TMYH13} in connection with the Jarzynski equation,
see also Appendix \ref{sec:PR1b}.

The above result that $\beta \Delta Q = \Delta S + O(\beta-\beta_0)^2$ can be viewed as a confirmation of the Clausius {\it identity}
in linear stochastic thermodynamics. The fact that the deviation to the Clausius identity is {\it non-negative} in the sense of Theorem \ref{T3}
can be made plausible in the following way. Consider a state change of an NLS with a slightly lower temperature than the heat bath, $\tau<\tau_0$,
consisting of two steps. In the first step there is a limited contact with the heat bath such that only the heat $\Delta_1 Q$ is flowing into the TLS
leading, in linear approximation, to an increase of its entropy  by $\Delta_1 S= \frac{\Delta_1 Q}{\tau}$. After this first step the system,
while being kept isolated, thermalizes and approximately assumes a Gibbs state with temperature $\tau_1$ such that $\tau<\tau_1<\tau_0$. This can be
reasonably expected if $N$  is large enough (or if $N=2$). In a second step there is another contact with the heat bath leading to
a further heat transfer of $\Delta_2 Q$ and, in linear approximation, to an increase of its entropy  by
$\Delta_2 S= \frac{\Delta_2 Q}{\tau_1}< \frac{\Delta_2 Q}{\tau}$.
The total heat transfer is $\Delta Q= \Delta_1 Q+ \Delta_2 Q $  and the total increase of entropy is $\Delta S= \Delta_1 S+ \Delta_2 S $
which is  {\it less} than $\frac{\Delta Q}{\tau}$. An analogous reasoning applies to the case of $\tau>\tau_0$ and a cooling of the NLS in two steps.
The ``first Clausius inequality"
$\beta \langle\Delta Q \rangle\ge \langle\Delta S\rangle$ thus reflects the fact that $\beta$ is the fixed initial inverse temperature of the NLS and possible changes of the NLS's temperature during the interaction with the heat bath are ignored in the term $\beta \langle\Delta Q\rangle$ but would be relevant for the term
$\langle\Delta S\rangle$.
On the other hand, the term $\beta \langle\Delta Q \rangle$ cannot be improved in a simple way,
because after the interaction with the heat bath,
the NLS may no longer be in a Gibbs state and thus has no temperature at all.\\

\begin{figure}[t]
\centering
\includegraphics[width=0.7\linewidth]{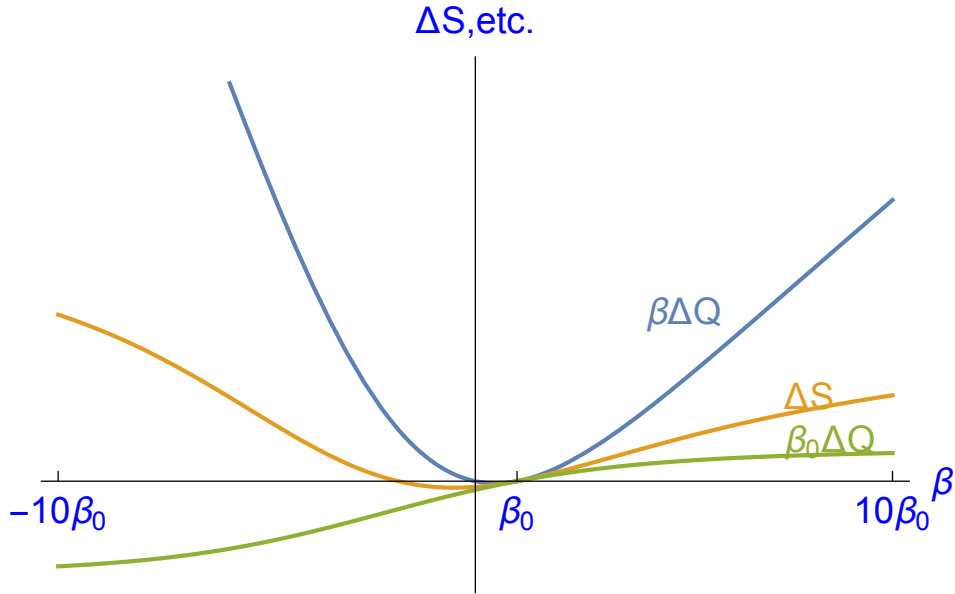}
	\caption{The same plot as in Figure \ref{FIGDES} but for the extended range of inverse temperatures $-10 \beta_0 \le \beta \le 10 \beta_0$
    of the NLS.
    Note that also in this extended range
    $\beta_0 \Delta Q\le \Delta S\le \beta \Delta Q$ holds in accordance with (\ref{theorem4}).
    For negative temperatures these ``Clausius inequalities" are less restrictive since here $\beta_0 \Delta Q$ and $\beta \Delta Q$
    have different signs. $\Delta S$ has a second zero for negative inverse temperatures that, in contrast to the first zero at $\beta=\beta_0$,
    does not correspond to a fixed point of the transition matrix $T$. For $\beta \rightarrow \pm \infty$ the Gibbs state probability $p(\beta)$
    is concentrated on the level with the lowest/highest energy, resp., and hence $\beta_0 \Delta Q$ as well es $\Delta S$ approach constant values.
    Therefore $\beta \Delta Q$ is asymptotically linear at these limits.
    }
\label{FIGDES1}
\end{figure}

Next we turn to a second Clausius inequality that can be obtained from the Jarzynski-type equation (\ref{Jar1})
by choosing $\widetilde{p}_m = q_m := \sum_n T_{mn}\,p_n$ for all $m\in{\mathcal N}$. This yields
\begin{equation}\label{Jar4}
 \left\langle \frac{p^{(0)}_n\,q_m}{p^{(0)}_m\,{p}_n}\right\rangle=1
 \;.
\end{equation}
As above we may invoke Jensen's inequality (JI) and the fact that $x\mapsto - \log x$ is a convex function:
\begin{eqnarray}
\label{Claus1}
  0&=& - \log 1\stackrel{(\ref{Jar4})}{=} - \log  \left\langle \frac{p^{(0)}_n\,q_m}{p^{(0)}_m\,{p}_n}\right\rangle\\
  \label{Claus2}
  &\stackrel{(JI)}{\le}& \left\langle
   -\log\frac{p_n^{(0)}}{d_n}  +\log\frac{p_m^{(0)}}{d_m} -\log\frac{q_m}{d_m} +\log\frac{p_n}{d_n}
   \right\rangle \\
   \label{Claus3}
   &\stackrel{(\ref{gibbsA1})}{=}& \left\langle
   \beta_0 E_n +\log Z_0 -\beta_0 E_m - \log Z_0 \right\rangle - \sum_{m}q_m\,\log \frac{q_m}{d_m}+ \sum_{n}p_n\,\log \frac{p_n}{d_n}\\
   \label{Claus4}
   &=& \langle\Delta S \rangle -\beta_0\, \langle\Delta Q\rangle
   \;,
\end{eqnarray}
where we have suitably expanded the fraction (\ref{Claus1}) with the factors $d_n$ and $d_m$.
Together with (\ref{RCI}) we have thus proven the following
\begin{theorem}\label{T4}
 \begin{equation}\label{theorem4}
   \beta_0\, \langle\Delta Q\rangle \le \langle \Delta S \rangle \le \beta\, \langle\Delta Q\rangle
   \;.
 \end{equation}
\end{theorem}

The second Clausius inequality $ \beta_0\, \langle\Delta Q\rangle \le \langle \Delta S\rangle $ has also be obtained
in \cite{Sa20}, Eq.~(4.4),  under the same conditions corresponding to our Assumption \ref{A1}
and using the ``monotonicity of the Kullback-Leibler (KL) divergence".
This proof is closely related to ours, since the said monotonicity is also a consequence of the general J-equation, see Appendix \ref{sec:GJE}.

Recall that according to Theorem \ref{T2} heat is always flowing from the hotter body to the colder one.
According to the first Clausius inequality (\ref{RCI}) the analogous statement for the entropy flow can only be shown in
the case $0\le\beta\le \beta_0$, i.~e., where the NLS has initially a larger temperature than the heat bath.
This follows since $\beta-\beta_0\le 0$ implies $\langle\Delta Q\rangle\le 0$ by Theorem \ref{T2} and hence
$\langle\Delta S\rangle \stackrel{(\ref{theorem4})}{\le}\beta\langle\Delta Q \rangle\le 0$.
The second Clausius inequality in (\ref{theorem4}) can now be used to extend the statement about the entropy flow
to the case of $\beta > \beta_0$, i.~e., where the NLS has initially a lower temperature than the heat bath.
In this case we always have $\langle\Delta Q\rangle \ge 0$ by Theorem \ref{T2} and hence
$\langle \Delta  S\rangle \stackrel{(\ref{theorem4})}{\ge} \beta_0\, \langle \Delta Q\rangle \ge 0$. We thus have proven the following
\begin{theorem}\label{T5}
Under the preceding conditions and for non-negative inverse temperatures of the NLS, i.~e., $\beta\ge 0$
there holds
\begin{equation}\label{theorem5}
  \left(\beta-\beta_0\right)\,\langle \Delta  S\rangle  \ge 0
  \;.
\end{equation}
\end{theorem}

\section{Bi-stochastic limit case}\label{sec:BL}
As remarked in Section \ref{sec:MR}, in the limit case $\beta_0=0$ and if $d_n=1$ for all $n\in{\mathcal N}$ we obtain the special case of a
bi-stochastic transition matrix $T$ satisfying (\ref{bistoch}). For $d_n\equiv 1$ the entropy (\ref{defSp}) can be identified with the
Shannon entropy \cite{S48}, up to the choice of units. Physically, this case can be realized by an NLS subject to external time-dependent forces
but not coupled to a heat bath. Although this special case is actually outside the thematic scope of this article, it will be instructive to
closer investigate it. The mathematics we used does not presuppose $\beta_0 \neq 0$ and hence this special case should be included in the preceding sections.
According to the mentioned physical realization of the bi-stochastic limit case we will refer to the random variable $\Delta E$
as ``work" and denote it by the variable $w$.

In particular, we consider the ``Clausius inequalities" (\ref{theorem4}) and re-write them as
\begin{equation}\label{CIBL}
  0 \le \langle \Delta S \rangle \le \beta \langle w \rangle
  \;.
\end{equation}
For $\beta\ge 0$ this implies $ \langle w \rangle \ge 0$, a result that could also have been derived from the usual
Jarzynski equation, see, e.g., \cite{S20}.

Another consequence of (\ref{CIBL}) is $\langle \Delta S \rangle \ge 0$
(in contrast to  $\langle\Delta S\rangle=0$ for classical adiabatic work processes).
This result can be independently proven as follows: Every bi-stochastic matrix $T$ can be written as a convex sum of permutational matrices.
This is the Birkhoff-von Neumann theorem, see \cite{B46,vN53}. The Shannon entropy is invariant under permutations,
but increases under a convex sum of probability distributions.  The latter is due to the concavity of the Shannon entropy, see, e.~g.,
\cite[ Ex.~$11.21$]{NC00}.

\section{Analytical example}\label{sec:AE}

As an example where the transition matrix $T$ can be exactly calculated we consider a single spin with spin quantum number $s=1$
coupled to a harmonic oscillator that serves as a heat bath. Hence we have a $3$-level system, ${\mathcal N}=\{1,0,-1\}$ and
$N=\left| {\mathcal N}\right|=3$. The total Hamiltonian is
\begin{equation}\label{totalH}
 H=H_1+H_2+H_{12}
 \;,
\end{equation}
where
\begin{eqnarray}\label{H1}
  H_1&=& \op{s}_z \otimes {\mathbbm 1}_\text{HO}\\
  \label{H2}
  H_2&=&  {\mathbbm 1}_\text{spin}\otimes \sum_{n=0}^\infty \left( n+\textstyle{\frac{1}{2}}\right)\left| n \right\rangle \left\langle n \right|\\
  \label{H12}
  H_{12}&=& \lambda \left(\op{s}^+\otimes A+\op{s}^-\otimes A^\ast \right)
  \;.
\end{eqnarray}
Here $\op{s}_x,\op{s}_y,\op{s}_z$ are the three spin operators and $\op{s}^\pm = \op{s}_x \pm{\sf i} \op{s}_y$
the two corresponding ladder operators. Similarly, $A$ and $A^\ast$ are the lowering and raising operators, resp.,
for the harmonic oscillator, such that $H_2= {\mathbbm 1}_\text{spin}\otimes\left(A^\ast\,A +\textstyle{\frac{1}{2}}{\mathbbm 1}_\text{HO}\right)$
and $\left( \left| n \right\rangle \right)_{n=0,1,2,\ldots}$ denotes the eigenbasis of $A^\ast\,A$. $\lambda$ is a real parameter.
Further, let $\left( \left| m\right\rangle \right)_{m=1,0,-1}$ be the eigenbasis of $\op{s}_z$ such that
$\left( \left| m,n\right\rangle\right)_{m\in{\mathcal N},n\in{\mathbbm N}}$ is an orthonormal basis of the total Hilbert space.

The Hamiltonian (\ref{totalH}-\ref{H12}) strongly resembles the Jaynes-Cummings model \cite{JC63}, which describes the interaction of a $2$-level system with a quantized radiation field. The extension to $3$-level systems has also been considered
\cite{GE90,AB95,TLV15}, but always assuming non-uniform level spacings and two radiation modes.

This system is analytically solvable since $H_{12}$ commutes with $H_1+H_2$. The  eigenspaces of $H_1+H_2$ are hence left invariant under $H$.
They are of the following form:
Either the singlet spanned by $\left| -1,0\right\rangle$, or the doublet spanned by  $\left| 0,0\right\rangle$ and $\left| -1,1\right\rangle$,
or an infinite number of triplets spanned by $\left| 1,n-1\right\rangle$, $\left| 0,n\right\rangle$ and $\left| -1,n+1\right\rangle$,
where $n=1,2,3,\ldots$.
Within the triplets $H$ has the form
\begin{equation}\label{tripH}
  H^{(n)}=\left(
\begin{array}{ccc}
 n+{\textstyle \frac{1}{2}} &  \lambda  \sqrt{2n} & 0 \\
 \lambda  \sqrt{2n} & n+{\textstyle\frac{1}{2}} &  \lambda  \sqrt{2n+2} \\
 0 &  \lambda  \sqrt{2n+2} & n+{\textstyle\frac{1}{2}} \\
\end{array}
\right)
\;,
\end{equation}
and the corresponding eigenvalues
\begin{equation}\label{tripeig}
  E^{(n)}_i\in\left\{n+{\textstyle\frac{1}{2}},n+{\textstyle\frac{1}{2}}\pm \lambda  \sqrt{4 n+2}\right\}
  \;.
\end{equation}

\begin{figure}[t]
\centering
\includegraphics[width=0.7\linewidth]{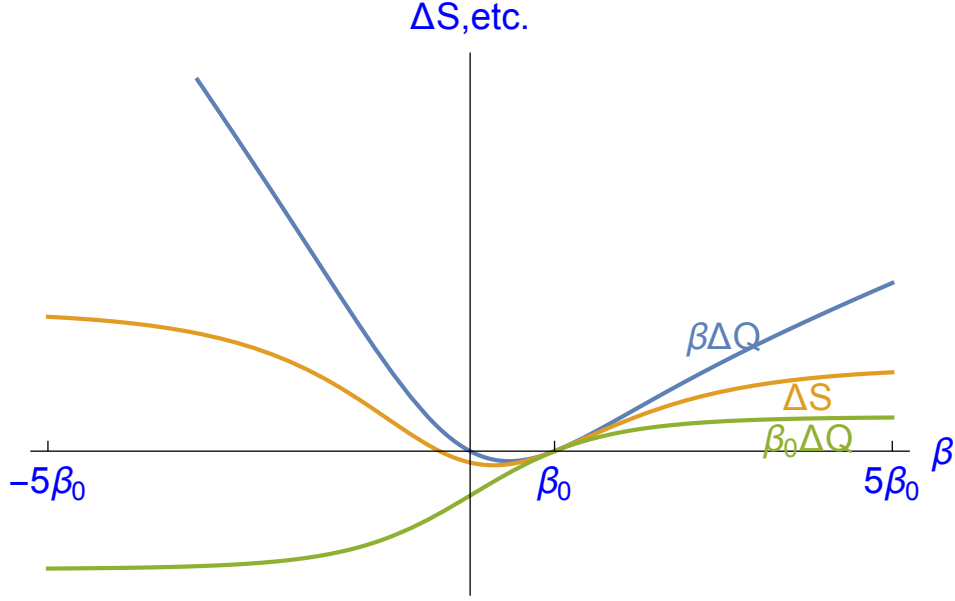}
	\caption{ Plot of the two ``Clausius heat terms" $\beta_0 \Delta Q$, $\beta \Delta Q$ and  the ``entropy increase"
$\Delta S$ as  functions of the inverse temperature $\beta$ of the $3$-level system considered in Section \ref{sec:AE}
and the corresponding transition matrix $T$ according to (\ref{T11}) - (\ref{T33}).
The range of the inverse temperature $\beta$ is chosen as $-5 \beta_0 \le \beta \le 5\beta_0$ and the calculations have been done for $\beta_0=1$.
It is remarkable that for $0<\beta< \beta_0$ the absolute value $|\Delta S|$ has its maximum not for $\beta=0$, as it is the case for $|\beta_0\Delta Q|$,
but for some positive inverse temperature of $\beta\approx 0.279896$.
}
\label{FIGDES2}
\end{figure}

Let $\pi$ denote a Gibbs state of the harmonic oscillator, such that
\begin{eqnarray}
\label{GSHO1}
  \pi_n &=& \frac{1}{Z}{\sf e}^{-\beta_0\left(n+\textstyle{\frac{1}{2}}\right)}\\
  Z &=& \sum_{n\in\mathbbm{N}}{\sf e}^{-\beta_0\left(n+\textstyle{\frac{1}{2}}\right)}=\frac{e^{-\frac{\beta _0}{2}}}{1-e^{-\beta _0}}
  \;.
\end{eqnarray}
We choose as an initial mixed state
\begin{equation}\label{inis}
 \rho=\sum_{mn}p_m \pi_n \left| mn\right\rangle\left\langle mn\right|
 \;,
 \end{equation}
 with an arbitrary probability distribution $p$. After the time $t$
 this state will evolve into
 \begin{equation}\label{rhop}
  \rho' = {\sf e}^{-{\sf i}H t}\,\rho\,{\sf e}^{{\sf i}H t}
  \;.
 \end{equation}

 To simplify the following calculation we will consider the {\it time average}
 \begin{equation}\label{rhopp}
  \rho'' =\overline{\rho'}=\overline{ {\sf e}^{-{\sf i}H t}\,\rho\,{\sf e}^{{\sf i}H t}}\
  \;.
 \end{equation}
 The time averaged probability of finally occupying the state $r\in{\mathcal N}$ will be
 \begin{equation}\label{qr}
   q_r=\sum_k \left\langle r k\right| \rho'' \left| r k\right\rangle =: \sum_m T_{rm}\,p_m
   \;.
 \end{equation}
 The latter equation follows since $q_r$ is a linear function of $p$ and defines the transition matrix $T$.
 From what has been said above it is clear that $T$ will be obtained by a summation over all eigenspaces of $H_1+H_2$.
 It proves to be independent of $\lambda$ due to time averaging.
 After some computer-algebraic calculations we obtain
 \begin{eqnarray}
\label{T11}
   T_{11} &=& \frac{1}{32} \left(e^{\beta _0}-1\right) \left(\frac{12}{e^{\beta _0}-1}+8
   e^{\frac{\beta _0}{2}} \coth ^{-1}\left(e^{\frac{\beta _0}{2}}\right)+3 e^{-\beta
   _0} \Phi \left(e^{-\beta _0},2,\frac{3}{2}\right)-8\right), \\
   \label{T12}
   T_{12} &=& \frac{1}{4} \left(1-2 \sinh \left(\frac{\beta _0}{2}\right) \tanh
   ^{-1}\left(e^{-\frac{\beta _0}{2}}\right)\right), \\
   \label{T13}
   T_{13}&=& \frac{3}{32} e^{-3 \beta _0} \left(4 e^{\beta _0}-\left(e^{\beta _0}-1\right) \Phi
   \left(e^{-\beta _0},2,\frac{3}{2}\right)\right),\\
   \label{T21}
   T_{21}&=&\frac{1}{4} e^{\beta _0} \left(1-2 \sinh \left(\frac{\beta _0}{2}\right) \tanh
   ^{-1}\left(e^{-\frac{\beta _0}{2}}\right)\right),\\
   \label{T22}
   T_{22}&=& \frac{1}{2},\\
   \label{T23}
   T_{23}&=&\frac{1}{4} e^{-\frac{3 \beta _0}{2}} \left(e^{\frac{\beta _0}{2}}+\left(e^{\beta
   _0}-1\right) \tanh ^{-1}\left(e^{-\frac{\beta _0}{2}}\right)\right),\\
   \label{T31}
   T_{31}&=&\frac{3}{32} e^{-\beta _0} \left(4 e^{\beta _0}-\left(e^{\beta _0}-1\right) \Phi
   \left(e^{-\beta _0},2,\frac{3}{2}\right)\right),\\
   \label{T32}
   T_{32}&=&\frac{1}{4} \left(2 \sinh \left(\frac{\beta _0}{2}\right) \tanh
   ^{-1}\left(e^{-\frac{\beta _0}{2}}\right)+1\right),\\
   \nonumber
   T_{33}&=&\frac{1}{32} e^{-3 \beta _0} \left(4 e^{2 \beta _0} \left(11 \sinh \left(\beta
   _0\right)+5 \cosh \left(\beta _0\right)-4 \sinh \left(\frac{\beta _0}{2}\right)
   \coth ^{-1}\left(e^{\frac{\beta _0}{2}}\right)-2\right)+3 \left(e^{\beta
   _0}-1\right) \Phi \left(e^{-\beta _0},2,\frac{3}{2}\right)\right)
   \;.\\
   \label{T33}&&
 \end{eqnarray}
 Here $\Phi(z,s,a):= \sum_{k\in{\mathbbm N}} \frac{z^k}{(k+a)^s}$  denotes the Lerch's transcendent, see
\cite[$\S\, 25.14$]{NIST}.
It can be shown that $T$ is a left stochastic matrix and leaves the probability distribution
\begin{equation}\label{Gibbs0}
 p^{(0)}=\frac{1}{e^{-\beta _0}+e^{\beta _0}+1}\left(e^{-\beta _0},1,e^{\beta _0}\right)
\end{equation}
 invariant that corresponds to a Gibbs state with inverse temperature $\beta_0$.
It follows that Assumption \ref{A1} is satisfied and hence the results derived in the  Sections
\ref{sec:MR} and \ref{sec:CI} hold for our example. We illustrate this by showing the three functions
$\beta \Delta Q$, $\Delta S$ and $\beta_0 \Delta Q$ for $-5 \beta_0 \le \beta \le 5 \beta_0$ in Figure \ref{FIGDES2}
satisfying the Clausius inequalities (\ref{theorem4}). For this example we make the following observation, which seems to be typical for
NLS coupled to a heat bath: If the $3$-level system is initially hotter than the heat bath, $0<\beta<\beta_0$,
the maximal heat transfer $|\Delta Q|$ results for $\beta \to 0$, as expected.
In contrast, the entropy transfer $|\Delta S|$ takes its maximum at a positive inverse temperature, see Figure \ref{FIGDES2}.

\begin{minipage}[t]{\textwidth}
\begin{center}\includegraphics[width=0.7\linewidth]{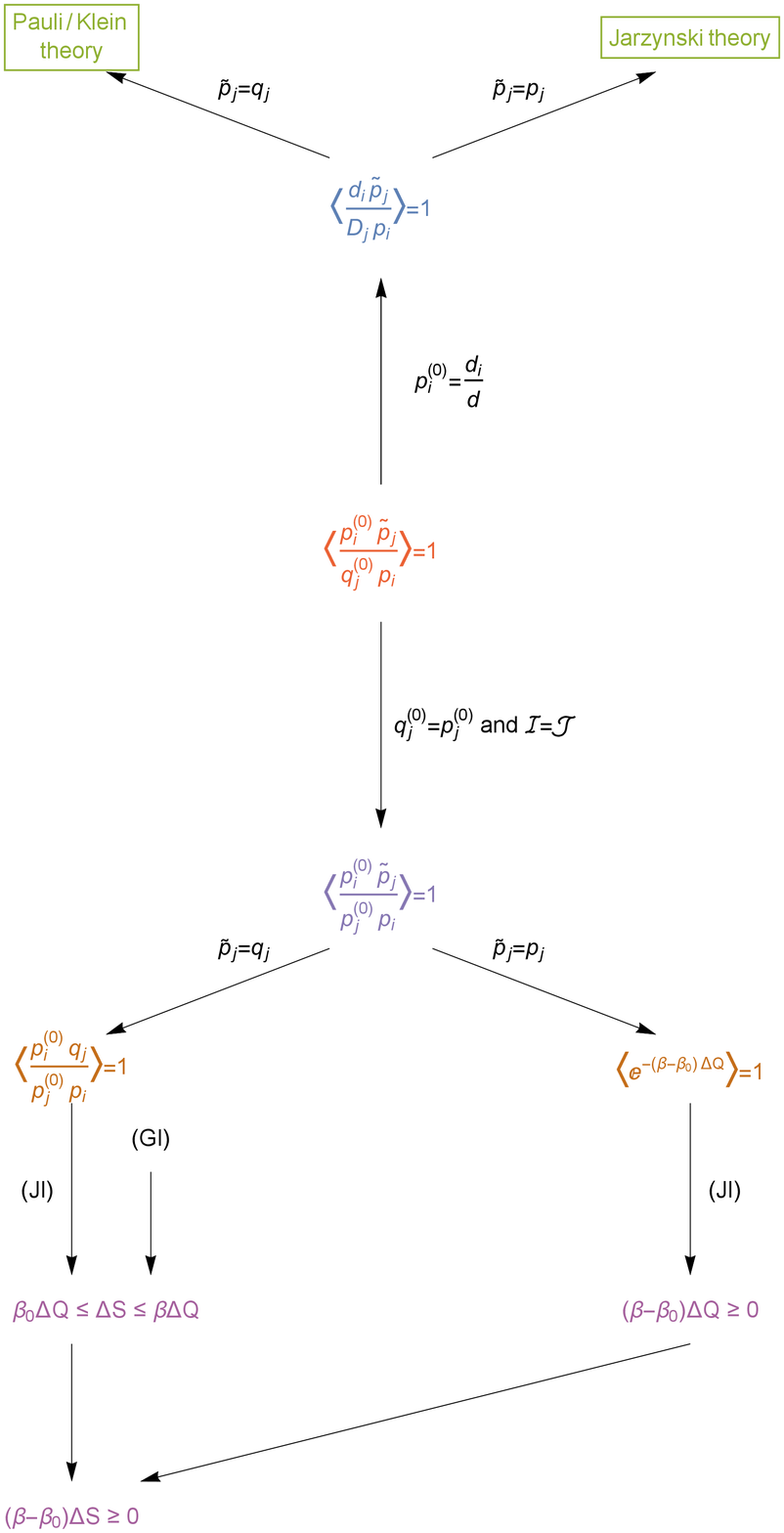}\label{FIGJG}
\end{center}
FIG 4: Schematic representation of various forms of J-equations and inequalities and their logical dependencies.
    Detailed explanations are given in Section \ref{sec:SO}.
\end{minipage}

\section{Summary and Outlook}
\label{sec:SO}

In this paper we have presented an approach to the time-honored problem of the $2^{nd}$ law for a finite quantum system
coupled to a heat bath. We have re-derived several known partial results, in part under weaker assumptions, and integrated them into a theory based on general J-equations. These resemble the famous Jarzynski equation and imply certain $2^{nd}$ law-like inequalities. It will be in order to provide
a general survey that shows their logical dependencies, see Figure 4.

The most general J-equation is (\ref{GJE1}), the central (red) equation of Figure 4. It holds for two
sequential measurements under rather general assumptions, see Theorem \ref{GJE}, and contains two undetermined probability
distributions $\widetilde{p}$ and $p^{(0)}$. There are two principal specialization paths that physically correspond to
``work processes" (upward direction in Figure 4) and ``heat processes" (downward direction in Figure 4).

For ``work processes" performed on closed systems that are only marginally touched in this paper
the transition matrix $T$ is bi-stochastic in the modified sense of Eq.~(\ref{bistochmod}).
This entails the J-equation (27) in \cite{SG20a} (the upper blue equation in Figure 4)
that can be further specialized according to the choice of $\widetilde{p}$. The usual Jarzynski equations are derived for the choice of
$\widetilde{p}=p$, whereas the alternative $\widetilde{p}=q:=T\,p$ leads to a scenario reminiscent of earlier  work of W.~Pauli
and F.~Klein, see \cite{SG20a,SG20b}.

By contrast, the ``heat processes" performed on systems without external forces but under contact with a heat bath
are characterized by $T\,p^{(0)}=p^{(0)}$ and lead to another special J-equation (\ref{Jar1}) (the lower gray equation in Figure 4)
that is of central importance for this work. Again, there are two further options. The choice $\widetilde{p}=p$ and the restriction to
probability distribution given by Gibbs states leads to (\ref{Jar3}) and, by means of Jensens's inequality (JI), to
the relation (\ref{JIT2}). The latter states that, on statistical average, heat always flows from the hot system to the cold bath and vice versa.

For the analogous statement (\ref{theorem5}) about the average flow of entropy we additionally required the two Clausius inequalities.
The first one, $\Delta S \le \beta \Delta Q$, is a simple consequence of the Gibbs inequality (GI).
The second one, $\beta_0 \Delta Q \le \Delta S$, follows via (JI) from the mentioned  J-equation (\ref{Jar1}) and the choice $\widetilde{p}=q:=T\,p$.

Moreover, we have presented an analytically solvable example illustrating our approach. It consists of a $3$-level system (a spin with $s=1$)
coupled to a harmonic oscillator. For this example our central Assumption \ref{A1}, saying that the transition matrix has a fixed point of Gibbs type,
is exactly satisfied.
It would be a task for the future to investigate the conditions under which this assumption holds exactly or approximately.

\appendix

\section{Proof of Proposition \ref{PCE}}
\label{sec:PCE}
First, we will calculate the  energy increase in the weak coupling limit:
\begin{equation}\label{PCE1}
  \Delta E = E(T\,p)-E(p) = \sum_{mn}T_{mn}\,p_n \,E_m-\sum_n p_n\,E_n
  \stackrel{(\ref{Tweak})}{=} \varepsilon\,\sum_{mn} t_{mn} \,p_n\,E_m +O\left(\varepsilon^2\right)
  \;.
\end{equation}
Recall that the (left) stochasticity of $T$ implies
\begin{equation}\label{sumtmn}
  \sum_m t_{mn}=0
  \;.
\end{equation}
Further,
\begin{equation}\label{dSdpm}
  \left.\frac{\partial }{\partial q_m}S(q)\right|_{q=p}= -\left.\frac{\partial }{\partial q_m}\sum_n q_n\,\log \frac{q_n}{d_n}\right|_{q=p} =
  -\left( 1+\log \frac{p_m}{d_m}\right)
  \;,
\end{equation}
and hence
\begin{eqnarray}
\label{dSde}
  \left.\varepsilon\,\frac{\partial }{\partial\varepsilon}\,S(q)\right|_{\varepsilon=0} &=&  \varepsilon\,\sum_m  \frac{\partial q_m}{\partial \varepsilon}
   \left. \frac{\partial }{\partial q_m}S(q)\right|_{q=p} \\
   & \stackrel{(\ref{Tweak},\ref{dSdpm})}{=}& - \varepsilon\,\sum_m \left( \sum_n t_{mn}\,p_n\right) \left(1+\log \frac{p_m}{d_m} \right)\\
    & \stackrel{(\ref{sumtmn})}{=}& - \varepsilon\,\sum_m \left( \sum_n t_{mn}\,p_n\right) \,\log \frac{p_m}{d_m} \\
    &\stackrel{(\ref{defpi})}{=}& \varepsilon\, \sum_{mn}   t_{mn}\,p_n\,\left( \beta E_m +\log Z\right)\\
     & \stackrel{(\ref{sumtmn})}{=}& \varepsilon\,\beta \,\sum_{mn}   t_{mn}\,p_n\,E_m \stackrel{(\ref{PCE1})}{=}\beta \,\Delta E +O\left(\varepsilon^2\right)
  \;.
\end{eqnarray}
Finally,
\begin{equation}\label{DeltaSeps}
  \Delta S = \left. \varepsilon\,\frac{\partial }{\partial\varepsilon}\,S(q)\right|_{\varepsilon=0}+O\left(\varepsilon^2\right)
  = \beta\,\Delta E +O\left(\varepsilon^2\right)
  \;,
\end{equation}
which completes the proof of  Proposition \ref{PCE}. \hfill$\Box$\\

\section{Proof of $a\ge 0$}
\label{sec:PR1}
We will present two proofs of the fact that the slope $a$ of $\beta_0 \langle\Delta Q\rangle$ considered as a function of $\beta$
will be non-negative at $\beta=\beta_0$.

\subsection{First proof}\label{sec:PR1a}

We write $a$ as double sum according to (\ref{betaDQ4}) and add the same sum but with $n$ and $m$ interchanged. This yields
\begin{equation}\label{a1}
 a= {\scriptsize \frac{\beta_0}{2}}\sum_{nm} \left(T_{nm}\,p^{(0)}_m\,E_m-T_{mn}\,p^{(0)}_n\,E_n \right) \left( E_m-E_n\right)
 \;.
\end{equation}
Using
\begin{equation}\label{Tijpj1}
 T_{nm}\,p^{(0)}_m = {\scriptsize \frac{1}{2}}\left( T_{nm}\,p^{(0)}_m + T_{mn}\,p^{(0)}_n  \right)+
                    {\scriptsize \frac{1}{2}}\left( T_{nm}\,p^{(0)}_m - T_{mn}\,p^{(0)}_n  \right)
 \;,
\end{equation}
and the analogous expression for  $T_{mn}\,p^{(0)}_n$ we obtain
\begin{eqnarray}
\label{Tijpj2}
  a &=& {\scriptsize \frac{\beta_0}{4}}\sum_{nm} \left(
  \left( T_{nm}\,p^{(0)}_m + T_{mn}\,p^{(0)}_n  \right)E_m\left(E_m-E_n \right)+
  \left( T_{nm}\,p^{(0)}_m - T_{mn}\,p^{(0)}_n  \right)E_m\left(E_m-E_n \right)\right.\\
  \nonumber
     && \left.-\left( T_{mn}\,p^{(0)}_n + T_{nm}\,p^{(0)}_m  \right)E_n\left(E_m-E_n \right)-
  \left( T_{mn}\,p^{(0)}_n - T_{nm}\,p^{(0)}_m  \right)E_n\left(E_m-E_n \right)\right)\\
  \label{Tijpj3}
  &=&{\scriptsize \frac{\beta_0}{4}}\sum_{nm} \left(
  \left( T_{nm}\,p^{(0)}_m + T_{mn}\,p^{(0)}_n  \right)\left(E_m-E_n \right)^2+
  \left( T_{nm}\,p^{(0)}_m - T_{mn}\,p^{(0)}_n  \right)\left(E_m^2-E_n^2 \right)
  \right)
\end{eqnarray}
The first term in (\ref{Tijpj3}) is non-negative and the second one vanishes according to
\begin{equation}\label{Tijpj4}
 \sum_{nm} \left( T_{nm}\,p^{(0)}_m - T_{mn}\,p^{(0)}_n  \right)E_m^2 =\sum_m \left(p^{(0)}_m \,E_m^2 - p^{(0)}_m\, E_m^2\right)=0
\;,
\end{equation}
and
\begin{equation}\label{Tijpj5}
 \sum_{nm} \left( T_{nm}\,p^{(0)}_m - T_{mn}\,p^{(0)}_n  \right)E_n^2 =\sum_n \left(p^{(0)}_n\,E_n^2 - p^{(0)}_n\,E_n^2\right)=0
\;.
\end{equation}
Here we have used that $T$ is left stochastic, see (\ref{Trans2}), and leaves $p^{(0)}$ fixed, see (\ref{TA1}),
according to Assumption \ref{A1}.
\hfill$\Box$\\

\subsection{Second proof}\label{sec:PR1b}

The second proof is based on the cumulant expansion for a random variable $X$, see, e.~g., \cite[$26.1.12$]{AS72},
\begin{equation}\label{cumu}
  \log \left\langle {\sf e}^{t X}\right\rangle =\sum_{n=1}^\infty \kappa_n \frac{t^n}{n!}=
  \left\langle X\right\rangle t+ \frac{1}{2}\sigma^2(X) t^2+\ldots
  \;,
\end{equation}
defining the \textit{cumulants} $\kappa_1=\left\langle X\right\rangle,\kappa_2=\sigma^2(X):= \left\langle X^2\right\rangle- \left\langle X\right\rangle^2, \kappa_3=\left\langle \left(X-\left\langle X\right\rangle\right)^3\right\rangle,\ldots $. In our case we set $X=\Delta Q$ and $t=-(\beta-\beta_0)$ and use Eq.~(\ref{Jar3}) to obtain
\begin{equation}\label{cumu1}
 0 = \log 1 \stackrel{(\ref{Jar3})}{=} \log \left\langle {\sf e}^{-(\beta-\beta_0) \Delta Q}\right\rangle
 \stackrel{(\ref{cumu})}{=} - \left\langle \Delta Q\right\rangle\,(\beta-\beta_0)+\frac{1}{2}\sigma^2(\Delta Q)(\beta-\beta_0)^2+O(\beta-\beta_0)^3
 \;.
\end{equation}
Hence
\begin{equation}\label{slope1}
  \beta_0 \,\left\langle \Delta Q\right\rangle = \frac{\beta_0}{2}\sigma^2(\Delta Q)(\beta-\beta_0)+O(\beta-\beta_0)^2
  \;.
\end{equation}
This proves that $a=\frac{\beta_0}{2}\sigma^2(\Delta Q)\ge 0$ since the variance $\sigma^2$ of every random variable is non-negative. \hfill$\Box$\\

\section{General J-equation}
\label{sec:GJE}

We will sketch the general probabilistic framework, analogous to that in  \cite{SG20a} and already used in Section \ref{sec:MR} for a special case.
It deals with two sequential measurements and the corresponding general J-equation.
We have two sets of outcomes, ${\mathcal I}$ for the first and ${\mathcal J}$ for the second measurement.
Further, there exists a probability distribution
\begin{equation}\label{ProbIJ}
 P:{\mathcal I}\times {\mathcal J}\rightarrow [0,1]
\end{equation}
of the form
\begin{equation}\label{PT1}
 P(i,j)=T_{ji}\,p_i,\quad i\in{\mathcal I} \mbox{ and } j\in{\mathcal J}
 \;,
\end{equation}
where $T$ is a (left) stochastic matrix and $p:{\mathcal I}\rightarrow [0,1]$ a probability distribution.
$P$ will be used to calculate expectation values $\langle Y\rangle$ for random variables $Y:{\mathcal I}\times {\mathcal J}\rightarrow \mathbbm{R}$.

For simplicity we assume that ${\mathcal I}$ and ${\mathcal J}$ are finite and $p_i>0$ for all $i\in{\mathcal I}$.
Let $\widetilde{p}:{\mathcal J}\rightarrow [0,1]$  and $p^{(0)}:{\mathcal I}\rightarrow (0,1)$ be two further
probability distributions and define
\begin{eqnarray}\label{defq1}
  q_j&:=& \sum_i T_{ji}\,p_i,\quad j\in {\mathcal J}\\
  \label{defq0}
  q_j^{(0)}&:=& \sum_i T_{ji}\,p_i^{(0)},\quad j\in {\mathcal J}
  \;.
\end{eqnarray}
Then the following holds:
\begin{theorem}\label{GJE}
(General J-equation)\\
 \begin{equation}\label{GJE1}
   \left\langle \frac{p_i^{(0)}\, \widetilde{p}_j}{q_j^{(0)}\,p_i}\right\rangle =1
   \;.
 \end{equation}
\end{theorem}

{\bf Proof}:
\begin{eqnarray}
\label{ProofGJE1}
   \left\langle \frac{p_i^{(0)}\, \widetilde{p}_j}{q_j^{(0)}\,p_i}\right\rangle &=& \sum_{ij}P(i,j)\,\frac{p_i^{(0)}\, \widetilde{p}_j}{q_j^{(0)}\,p_i} \\
   \label{ProofGJE2}
   &\stackrel{(\ref{PT1})}{=}& \sum_{ij}T_{ji}\,\xout{p}_i\,\frac{p_i^{(0)}\, \widetilde{p}_j}{q_j^{(0)}\,\xout{p}_i}\\
   \label{ProofGJE3}
     &\stackrel{(\ref{defq0})}{=}&\sum_{j}\xout{q}_j^{(0)}\,\frac{\widetilde{p}_j}{\xout{q}_j^{(0)}}=\sum_j \widetilde{p}_j=1
     \;.
\end{eqnarray}
\hfill$\Box$\\

The general J-equation (\ref{GJE1}) is more of a template that can be used to generate
further equations by choosing a special form of the general probability distribution $\widetilde{p}$.
Note that it is not required that the probability distribution $p^{(0)}$ is invariant under $T$; if this is the case
and, moreover, ${\mathcal I}={\mathcal J}\equiv {\mathcal N}$ then the special form of Eq.~(\ref{Jar1}) results.

The J-equation (27) in \cite{SG20a} follows for the choice of $p^{(0)}_i=\frac{d(i)}{d}$ for all $i\in{\mathcal I}$,
where $d:=\sum_i d(i)$ that leads to a modified bi-stochasticity of $T$ in the sense of  \cite[Eq.~(25)]{SG20a}.

As for every Jarzynski-type equation the application of Jensen's inequality (JI) using the concave function $x\mapsto \log x$
yields a $2^{nd}$ law-like inequality. In the case of (\ref{GJE1}) this inequality turns out to be equivalent to
the ``monotonicity of the Kullback-Leibler (KL) divergence", see \cite{Sa20}, if we set $\widetilde{p}_j =q_j$ for all $j\in{\mathcal J}$.
This will be shown in the following.

\begin{eqnarray}
\label{KL1}
  0 &=& \log 1\stackrel{(\ref{GJE1})}{=} \log \left\langle \frac{p_i^{(0)}\,q_j}{q_j^{(0)}\,p_i}\right\rangle \\
  \label{KL2}
  &\stackrel{(JI)}{\ge}& \left\langle \log\, \frac{p_i^{(0)}\, q_j}{q_j^{(0)}\,p_i}\right\rangle
  = \left(\left\langle \log q_j\right\rangle - \left\langle \log q_j^{(0)}\right\rangle\right) -
  \left(\left\langle \log p_i\right\rangle - \left\langle \log p_i^{(0)} \right\rangle\right)
 \;.
\end{eqnarray}
We further obtain
\begin{equation}\label{logq1}
 \left\langle \log q_j \right\rangle = \sum_{ij}T_{ji}\,p_i\,\log q_j = \sum_j q_j\,\log q_j
 \;,
\end{equation}
and
\begin{equation}\label{logq0}
 \left\langle \log q_j^{(0)} \right\rangle = \sum_{ij}T_{ji}\,p_i\,\log q_j^{(0)} = \sum_j q_j\,\log q_j^{(0)}
 \;,
\end{equation}
hence
\begin{equation}\label{relentr}
 \left\langle \log q_j\right\rangle - \left\langle \log q_j^{(0)}\right\rangle= \sum_j q_j \log\frac{q_j}{q_j^{(0)}}
 =S\left( q \| q^{(0)}\right)
 \;.
\end{equation}
Analogously,
\begin{equation}\label{relentr0}
 \left\langle \log p_i\right\rangle - \left\langle \log p_i^{(0)}\right\rangle = \sum_i p_i \log\frac{p_i}{p_i^{(0)}}
 =S\left( p \| p^{(0)}\right)
 \;,
\end{equation}
and hence (\ref{KL2}) is equivalent to the monotonicity of the KL-divergence written as
\begin{equation}\label{DKL}
S\left( q \| q^{(0)}\right) \le  S\left( p \| p^{(0)}\right)
\;.
\end{equation}

As mentioned in the Introduction the general J-equation belongs to the framework of stochastic thermodynamics
and is as such neither quantum nor classical.
Nevertheless, it will be instructive to sketch realisations of the framework in these two domains.

We begin with the quantum domain. Consider a finite-dimensional system $\Sigma_1$ initially described by a statistical operator with spectral decomposition
$\rho_1 =\sum_i x_i P_i$ such that
\begin{equation}\label{trq1}
  1 = \mbox{Tr} \rho_1= \sum_i x_i \mbox{Tr} P_i =: \sum_i x_i\,d_i =: \sum_i p_i
  \;.
\end{equation}
A first projective measurement corresponding to the complete family of mutually orthogonal projections $\left( P_i\right)_{i\in{\mathcal I}}$
leaves $\rho_1$ invariant.
The system is then coupled  to some auxiliary system $\Sigma_2$ with initially mixed state $\rho_2$ and the total system undergoes
a finite time evolution described by some unitary operator $U$ defined in the total Hilbert space.
Finally, a quantum measurement at $\Sigma_1$  is performed corresponding to a complete family of mutually orthogonal
projections $\left( Q_j\right)_{j\in{\mathcal J}}$. The probability $q_j$ of the outcome $j\in{\mathcal J}$ of the final measurement is
given by
\begin{equation}\label{probqj}
 q_j= \mbox{Tr} \left(\left(Q_j\otimes \mathbbm{1} \right)U\left( \rho_1\otimes \rho_2\right) U^\ast \right)=
 \sum_i p_i \mbox{Tr} \left(\left(Q_j\otimes \mathbbm{1} \right)U\left(\frac{P_i}{d_i}\otimes \rho_2\right) U^\ast \right)
 =:\sum_i T_{ji}\,p_i
 \;.
\end{equation}
It is straightforward to check that the matrix $T$ defined by (\ref{probqj}) is (left) stochastic
using $\sum_j Q_j=\mathbbm{1}$.\\

We note that it is not necessary to take the total initial state as a product state $\rho_1\otimes \rho_2$,
although this is usually assumed in the literature on the $2^{nd}$ law, e.~g., in  \cite[p.~$113$ ]{SL78}, or
\cite[p.~$230602-3$]{JW04}, and also in this paper, see Section \ref{sec:MR}.
But if we assume an arbitrary total initial state $\rho$ and perform the first projective measurement
according to the family $\left( P_i\otimes {\mathbbm 1}\right)_{i\in{\mathcal I}}$,
then the resulting state will be $\rho' = \sum_{i\in{\mathcal I}} \left( P_i\otimes {\mathbbm 1}\right)\,\rho\,\left( P_i\otimes {\mathbbm 1}\right)$,
which is not entangled but may, nevertheless, have some ``classical" correlation. It yields the initial probabilities
$p_i=\mbox{Tr } \left( P_i\otimes {\mathbbm 1}\right)\,\rho$ and
\begin{equation}\label{qjcorr}
 q_j= \mbox{Tr} \left(\left(Q_j\otimes \mathbbm{1} \right)
 U\left(\sum_i p_i
 \frac{ \left( P_i\otimes {\mathbbm 1}\right)\,\rho\,\left( P_i\otimes {\mathbbm 1}\right)}
 {\mbox{Tr } \left( P_i\otimes {\mathbbm 1}\right)\,\rho}\right) U^\ast \right)
 =:\sum_i T_{ji}\,p_i
 \;,
\end{equation}
analogously to (\ref{probqj}).\\

In the classical case we work with a phase space $\left( {\mathcal P},d\mu\right)$ and an initial probability density $\rho$
satisfying
\begin{equation}\label{normrho}
  \int_{\mathcal P}\rho\,d\mu =1
  \;.
\end{equation}
The phase space is decomposed according to the finite partition ${\mathcal P}=\biguplus_{i\in{\mathcal I}} {\mathcal P}_i$
and $\rho$ is correspondingly written as
\begin{equation}\label{decrho}
\rho=\sum_i \chi_i\,\rho=\sum_i p_i \, \rho_i
\;,
\end{equation}
where
\begin{equation}\label{defrhoi}
p_i:= \int_{{\mathcal P}_i}\rho\,d\mu\;,\quad
 \rho_i :=  \frac{\chi_i\,\rho}{p_i}
 \;,
\end{equation}
and $\chi_i$ is the characteristic function of ${\mathcal P}_i$ for all $i\in{\mathcal I}$.
The time evolution of the classical system is described by a measure-preserving map $U:{\mathcal P} \rightarrow{\mathcal P}$,
such that $\rho$ will be transformed into $\rho' = \rho \circ U^{-1}$. Finally, a discrete  measurement is performed
according to another finite partition ${\mathcal P}=\biguplus_{j\in{\mathcal J}} {\mathcal Q}_j$. The probability $q_j$
of finding the system in the subset ${\mathcal Q}_j$ is given by
\begin{equation}\label{probQj}
 q_j= \int_{{\mathcal Q}_j} \rho'\,d\mu= \sum_i p_i \, \int_{{\mathcal Q}_j} \rho_i\circ U^{-1}\,d\mu=:\sum_i T_{ji}\,p_i
 \;.
\end{equation}
It is straightforward to check that the matrix $T$ defined by (\ref{probQj}) is (left) stochastic.

\section*{Acknowledgment}
This work was funded by the Deutsche Forschungsgemeinschaft (DFG) Grants No.~SCHN 615/25-1 and GE 1657/3-1.
We sincerely thank the members of the DFG Research Unit FOR2692 for fruitful discussions.


\end{document}